# Channeling of a sub-angstrom electron beam in a crystal mapped to two-dimensional molecular orbitals

Robert Hovden,* Huolin L. Xin,** and David A. Muller*
* School of Applied and Engineering Physics, Cornell University, Ithaca, NY 14853
** Department of Physics, Cornell University, Ithaca, NY 14853
(Dated: Apr 28, 2011)

The propagation of high-energy electrons in crystals is in general a complicated multiple scattering problem. However, along high-symmetry zone axes the problem can be mapped to the time evolution of a two-dimensional (2D) molecular system. Each projected atomic column can be approximated by the potential of a 2D screened hydrogenic atom. When two columns are in close proximity, their bound states overlap and form analogs to molecular orbitals. For sub-angstrom electron beams, excitation of anti-symmetric orbitals can result in the failure of the simple incoherent imaging approximation. As a result, the standard resolution test and the one-to-one correspondence of atomic positions of a crystal imaged along a zone-axis with closely spaced projected columns ("dumbbells") can fail dramatically at finite and realistic sample thicknesses. This is demonstrated experimentally in high angle annular dark field scanning transmission electron microscope (HAADF STEM) images of [211]–oriented Si showing an apparent inter-column spacing of 1.28(±.09) Å, over 64% larger than the actual 0.78 Å spacing. Furthermore, the apparent spacing can be tuned with sample thickness and probe size to produce a larger, smaller, or even the actual spacing under conditions when the peaks of two adjacent Si columns should not even have been resolved given the electron probe size.

I. INTRODUCTION

The propagation of a high-energy electron beam through a three-dimensional (3D) periodic potential can be mapped to the time-evolution of a wavepacket in an array of 2D projected potentials [1-4]. Previous real-space and Bloch-wave [5,6] models are effective descriptions of well-separated atomic columns, but the simple image interpretations that can be extracted from these models are recognized to fail for the closely-spaced features accessible to the new generation of aberration-corrected electron microscopes [7,8]. The form of the failure has not yet been considered—here we show it leads to a new scattering regime with very real consequences for image interpretation. We present a tight-binding model for swift electron propagation in crystals where paired atomic columns can be treated in analogy to a two-dimensional (2D) hydrogenic molecule. In isolation, each column produces a transverse set of bound and continuous states, resembling that of a 2D hydrogenic atom. When two atomic columns are in close proximity, the overlapping bound states give rise to symmetric and anti-symmetric orbitals. When a scanning transmission electron microscope (STEM) is used to probe a crystal structure, the excitation of anti-symmetric orbitals can make paired atomic columns separated by less than the probe width appear to be incorrectly resolved, placing images of the projected columns at unphysical locations. Prior studies of overlapping orbitals [9,10] had considered a regime where the atomic columns were separated by more than an Angstrom and the potential effects of overlap and coupling are exponentially weaker. Consequently, in P. Geuens & Van Dyke [11] the authors concluded coupling between columns can be neglected. While their conclusions were appropriate for the older generation of lower-resolution microscopes where the coupling between sufficiently widely-spaced columns can be neglected, for more modern instruments capable of forming sub-Angstrom sized beams, this is no longer the case and the resulting distortions often prevent a simple interpretation of images and may raise doubts about real-space resolution measurements or atomic displacements such as ferroelectric distortions in a material.

A major challenge in measuring the experimental probe profile of a sub-angstrom electron wavepacket is the lack of good test objects. There are no bond lengths shorter than 0.5 Å. Instead, most resolution tests rely on viewing a thin section of a crystal along a zone axis such as silicon [11*n*] (*n* = 2, 4,..) where the projected distance between neighboring atom columns can be shrunk incrementally [12]. In projection, the resulting closely spaced pair of atomic atoms is referred to as a dumbbell due to the shape of its image when the pair is just resolved. An implicit assumption in this test is that the shape of the probe wave function is not altered by the atomic potentials. However incident swift electrons are known to channel along the positively charged screened potential of an atomic column's nuclei [13-16].

Here, we explain how electron channeling can result in image artifacts, a dramatic failure of standard resolution tests, and confound attempts to measure atomic displacements in interfaces and ferroelectric thin films. Experimentally, these effects are shown in a scanning transmission electron microscope (STEM) image (Fig. 1), where the Si [211] paired columns are located 68%[17] further away than the actual inter-column spacing. Considering the resolution degradation from chromatic aberration and the incoherent source size, the paired column peaks of Si [211] are not expected to be resolved in this micrograph. This is in agreement with the fact that no information out to 0.78 Å (i.e. 444) is found in the diffractogram (Fig. 1, also seen in Supp. Fig. 1)[18]. Previous work has emphasized real-space resolution measurements since information in the diffractogram can lead to false positives[17, 19, 20]. However, simply checking for the appearance of dumbbell structure in an image could lead many to overestimate the resolution of their STEM. A combination of Fourier analysis, peak-to-peak measurement, ideally over a range of sample thicknesses and backed by multislice simulation is needed to best verify resolution. These shifts are also much larger than many atomic displacement expected near grain boundaries or interfaces.

II. METHODS

Understanding the dynamic scattering of high-energy electrons in crystals is often tackled by solving Bethe's equation using Bloch-waves [21, 22]. Plane-wave Bloch *s*-states are a truncated (and slowly converging) Fourier expansion of the atomic columnar *s*-states considered here. When a small number of plane-wave Bloch beams is selected to keep the model analytically tractable and interpretable, the Bloch s-state has a very different shape and spatial extent from the columnar s-state. The Bloch convergence is worst for the on-column features, which are those that contribute most to the ADF-STEM intensity. Consequently, describing a highly localized, channeled electron wave may not be best suited for the extended, periodic plane-wave Bloch basis that relies on the extended translational symmetry of crystal in the transverse direction, a symmetry which is broken at interfaces, grain boundaries, and defects. Nevertheless, when fully converged (which scales as the cube of the number of beams, $O(N^3)$ - as shown by Fig 6.2 of ref [23]), it does capture many of properties of isolated columns, a result best understood by again appealing to atomic models. Here we will focus on the direct calculation of the real-space s-states in a local basis, an approach that scales linearly with the number of atomic columns.

Numerical solutions can also be obtained with $O(N^2 \log N)$ scaling by a real space, Green's-function multislice formulation incorporating a frozen phonon model [24, 25], and we will use this to check our simpler, analytic models. Forward-propagating fast electrons of constant velocity in materials can be described by a scalar-relativistic-corrected time-dependent Schrodinger equation with the time *t* replaced by the position in the forward direction, *z*:

$$\frac{\partial \psi}{\partial z} \quad [\frac{i\lambda}{4\pi}\nabla^2_{xy} + \frac{2mei\lambda}{4\pi\hbar^2}V(x,y,z)]\psi(x,y,z) \qquad (1)$$

where x,y,z are Cartesian spatial coordinates, *e* is the electron charge, and *m* and λ are the relativistic electron mass and wavelength respectively [23]. By numerical application of this method, Fig. 2a shows the free propagation of a 100 keV electron beam with a semi-convergent angle of 33 mrad. The electron beam converges at the focal plane and then diverges rapidly. However, when a column of Si atoms is present and aligned with the incident beam direction (Fig. 2b), a significant fraction of the electrons are attracted to and channel along the column.

In an *s*-state model, the electron channeling is assumed to be predominantly from the excitation and propagation of the *1s* transverse bound state of a projected atomic column, although more generally a larger family of bound and unbound states needs to be considered [26]. In the first order approximation, a fast moving incident electron (60-300 keV) experiences the average potential along its direction of motion. When a crystal is projected down a high symmetry zone axis, atoms aligned along the zone can be approximated as uniform columns of charge. The Ewald sphere is treated as a flat surface and the excitations on the high order Laue Zones are ignored [1,2]. At lower energies and larger angles, this approximation can break down [27]. In this approximation, the potential is z-independent and radially symmetric. The propagating wave function can be written as a linear combination of transverse eigenstates of a 2D time-independent Schrodinger's Equation,

$$\psi(\rho,\theta,z) \quad \sum_{n,\ell} c^{(n,\ell)} \frac{\phi_{n,\ell}(\rho)}{\rho^{1/2}} e^{i\ell\theta} \exp(-i2\pi \frac{me\lambda E_t^{(n,\ell)}}{h^2} z) \qquad (2)$$

where $E_t$ is transverse eigenenergy of the particle and $\rho$, $\theta$, are the planar transverse polar coordinates. Each eigenstate is indexed by quantum numbers $n$, $\ell$ and weighted by the overlap coefficient with the initial probe wavefunction, $c^{(n,\ell)} \quad <\Psi(\rho,\theta,0)|\psi_t^{(n,\ell)}(\rho,\theta)>$. As discussed by Berry [1], the 2D radial Schrödinger equation differs from the 3D radial equation by containing an ($\ell^2$ - ¼) term in the centripetal potential instead of the familiar $\ell(\ell+1)$, with $\ell$ being the angular momentum quantum number. In contrast to the 3D case, the radial wave equation acquires an attractive centripetal potential when $\ell = 0$ (s-states). The transverse bound states (i.e. columnar orbitals) of each isolated atomic column are analogous to that of a 2D hydrogen atom [28]—with only the s-states having non-zero values at the origin.

This formalism permits both the bound atomic-like states and unbound states that can be written as a linear combination of Bessel functions (Laurent series) [1]. The unbound states of the columnar potential can be converged more efficiently by avoiding the rapid oscillatory behavior required near atomic nuclei by constructing waves orthogonalized to the bound eigenstates [29]. The resulting pseudopotential then may be sufficiently weak enough to justify a weakly-scattering calculation of unbound states. For any incident electron beam, the wavefunction propagation is determined by matching the appropriate phase and amplitudes at the entrance surface to the bound eigenstates, which propagate according to Eq. 2, while the remaining uncoupled states propagate as unbound, weakly-scattered waves in the crystal:

$$\psi_{fr}(\rho,\theta,z) \quad [\Psi(\rho,\theta,0)-\sum_{n,\ell}c^{(n,\ell)}\psi_t^{(n,\ell)}(\rho,\theta)]\exp(\frac{-i2\pi z}{\lambda}) \qquad (3)$$

The higher the beam energy, and thinner the crystal, the better this approximation becomes, but the pseudopotential experienced by the orthogonalized states will always be weaker than the original potential.

III. RESULTS

A. Si [211] as a Two-Level System

When two projected potentials are brought together, the time-independent Schrodinger equation can be approximately solved using a linear combination of the columnar orbitals [9, 11]. In this tight binding approach, a two-level system arises from the overlap of two closely-spaced columnar orbitals and the energy splitting of the resulting bonding and anti-bonding states (Fig. 3,4). Such a system arises in the dumbbell structure of Si along the [211] zone axes.

For a 100 keV electron, a single column of silicon atoms along the [211] direction only permits a single *1s* bound state, which is broader than 1 Å—a severe issue for sub-angstrom imaging. However, as two atomic columns are brought together, the bound states overlap and give rise to a two-level system comprising of a bonding and anti-bonding state. This has a pronounced impact on the electron propagation as shown in Figure 2c, which implies a signal delocalization as the electron beats between two columns. The behavior changes little when the full lattice is added (Fig. 2d) except to introduce a slightly faster damping envelope, indicating that the local bound states dominate the scattering and propagation behavior.

The two-level molecular system provides a transparent understanding for the unintuitive 'jumping' of a channeled beam between adjacent columns: Down the silicon [211] zone axis, the bonding and anti-bonding states made from two *1s* columnar orbitals located on their parent columns are shown in Fig. 3 for a 100 keV probing electron. The column pair has a 0.78 Å inter-column spacing and atomic column density of 1.50 atoms / nm. We used the screened atomic potentials tabulated by Kirkland [23] and solved the *1s* bound state—the only bound state for this system—numerically using the Numerov method [30]. When comparing the two states of the columnar pair, the anti-bonding state has a central node, an increased electron density in the tails outside the columnar pair, as well as a 28% larger on-column magnitude than the bonding state. The energies of the bonding and anti-bonding states are respectively -18.28 and -3.37 eV for a 100 keV electron and -21.89 and -9.15 eV for a 300 keV electron (two beam energies typical to current aberration-corrected microscopes). The shape of the bound s- states changes with beam energy due to scalar relativistic effects. Length contraction causes faster electrons to experience deeper potential wells, resulting in bound eigenstates with lower energies and faster radially decaying *1s* states (Fig. 3a). With less overlap in adjacent column's s-states, there is a smaller energy splitting between eigenstates (Fig. 3b).

As the bonding ($\psi_t^{(b)}$) and anti-bonding ($\psi_t^{(a)}$) states propagate with periods inversely proportional to their energy, they constructively and destructively interfere, resulting in a beating of wave intensity between the two columns. The period of beating between columns is inversely proportionally to the difference in their energies ($h^2/me\lambda(E_t^{(a)} - E_t^{(b)})$). The total wave function intensity of the two-level system is described by: $I \propto |c^{(a)}\psi_t^{(a)} + c^{(b)}\psi_t^{(b)}| \propto 1 + 2c^{(a)}c^{(b)}\cos(\frac{2\pi me\lambda}{h^2}(E_t^{(a)} - E_t^{(b)}))$ . Fourier analysis of the oscillations along each of the columns simulated by the multislice method (Supp. Fig. 1) shows a single strong peak with a wavelength of 45.51 and 75.85 nm for the 100 and 300 keV electrons, which matches within 0.3% the eigenenergy difference calculated by the tight-binding model[18]. Plotting the intensity of the two-level system with the addition of unbound states (Fig. 2e), we see that the periodicity of the channeled electron's wavefunction matches well with the multislice simulation.

B. Annular dark-field signals from the bonding and anti-bonding states

In annular dark-field STEM, the image is formed by scanning the beam across the sample and incoherently collecting the electrons that scatter to an annular dark-field (ADF) detector. Because the local scattering potential of atoms are strongly peaked at the atomic nuclei, the ADF signal is approximately proportional to the integrated probe intensity along atomic columns [13]. An interesting scattering regime arises for depths beyond the microscopes depth of the focus, where unbound components of the probe are sufficiently spread out and only contribute to the background level of the ADF signal. As the specimen thickness increases beyond the depth of focus, only the channeled beam intensity remains and plays a dominant role (Supp. Fig. 2)[18]. The excitation coefficient of each state, $c^{(j)}$, is given by the inner product of the probe at the entrance surface and that eigenstate. For the channeled electron beam along two adjacent and equivalent atomic columns, the contributed ADF signal at a given depth is approximately proportional to the change in beam intensity along each atomic column positioned at $r_1$ and $r_2$:

$$dI(\vec{\rho},z)/dz \propto \left|c^{(a)}\psi_t^{(a)}(\vec{r}_1,z)+c^{(b)}\psi_t^{(b)}(\vec{r}_1,z)\right|^2 + \left|c^{(a)}\psi_t^{(a)}(\vec{r}_2,z)+c^{(b)}\psi_t^{(b)}(\vec{r}_2,z)\right|^2 \quad (4)$$

where ρ is the incident beam position, and $z$ is depth. Expanding the terms:

$$dI(\vec{\rho},z)/dz \propto \left|c^{(a)}\right|^2\left(\left|\psi_t^{(a)}(\vec{r}_1,z)\right|^2+\left|\psi_t^{(a)}(\vec{r}_2,z)\right|^2\right)+\left|c^{(b)}\right|^2\left(\left|\psi_t^{(b)}(\vec{r}_1,z)\right|^2+\left|\psi_t^{(b)}(\vec{r}_2,z)\right|^2\right)$$
$$+(c^{(a)*}c^{(b)}+c^{(a)}c^{(b)*})(\psi_t^{(a)}(\vec{r}_1,z)\psi_t^{(a)}(\vec{r}_2,z)+\psi_t^{(a)}(\vec{r}_1,z)\psi_t^{(a)}(\vec{r}_2,z)) \quad (5)$$

For two adjacent, equivalent columns containing a 2-level system the bonding and anti-bonding states are symmetric and anti-symmetric such that:

$$\psi_t^{(b)}(\vec{r}_1,z) \quad \psi_t^{(b)}(\vec{r}_2,z) \text{ and } \psi_t^{(a)}(\vec{r}_1,z) \quad -\psi_t^{(a)}(\vec{r}_2,z) \quad (6)$$

and the cross terms cancel out,

$$dI(\vec{\rho},z)/dz \propto \left|c^{(a)}\right|^2\left|\psi_t^{(a)}(\vec{r}_1,z)\right|^2 + \left|c^{(b)}\right|^2\left|\psi_t^{(b)}(\vec{r}_1,z)\right|^2 \quad (7)$$

Further simplifying the expression, we can drop the z-dependence, exp($-i2\pi me\lambda Ez/h^2$), in terms that have squared magnitude:

$$dI(\vec{\rho},z)/dz \propto \left|c^{(a)}\right|^2\left|\psi_t^{(a)}(\vec{r}_1)\right|^2 + \left|c^{(b)}\right|^2\left|\psi_t^{(b)}(\vec{r}_1)\right|^2 \quad (8)$$

This constant scattering rate that does not vary with thickness is very different to the enhancement and depletion seen at the entrance surface. For realistically thick specimens where channeling behavior dominates, the ADF signal depends on the excitation coefficients and the on-column intensity of the bound Eigenstates. When the excitation coefficient magnitudes change very slowly with $z$, the signal, $I(z)$, from the channeled beam is approximately linear with thickness. As a measure of the variation of the excitation coefficients for a typical case of a 100 keV beam propagating in Si [211], the on-column intensity drops roughly 15% from 50 to 100 nm (Supp. Fig. 2,3)[18].

C. Failure of The Linear Imaging Model

The Si [211] anti-bonding state, with a 28% larger on-column probability density than the bonding state, scatters more strongly to high angles. We found that the probe positions where the maximum excitation of the Si [211] anti-bonding state occur deviate from the positions where the atomic columns are actually located (Fig. 5). The ADF signal from excitation of the anti-bonding state will have an inter-peak spacing of 0.92 Å (17% larger than 0.78 Å) for a 100 keV probe ($\alpha_{max}$ = 33 mrad, aberration free) focused on the entrance surface (Fig. 5a). If the probe is focused 12 nm into the sample, the excitation coefficients change and there is a dramatic increase in maximum inter-peak spacing— 1.48 Å or 89% increase (Fig. 5b, Fig. 6). Additionally, there will be little to no excitation of the anti-bonding state when the probe is positioned near the node of the anti-symmetric state. As a result, signal contributions from the scattered anti-bonding states cause closely-spaced dumbbells to appear wider than the actual spacing and with an enhanced inter-column contrast. While the model provides an upper bound to the observed spacing, the exact value is a sensitive function of the probe shape, defocus, and sample thickness.

The increased spacing of adjacent columns in a HAADF image may seem counterintuitive. A simple linear imaging model, where images are assumed to be the scattering potential convolved with the intensity of the unperturbed wave function [23], would result in two overlapping airy disks only capable of producing a smaller peak-to-peak spacing with less contrast [12]. However, the simple linear imaging model is seen to fail at thicknesses greater than ~10 nm, typical for current imaging conditions (Fig. 1, 7). Atoms can appear resolved (but at incorrect locations) under unresolvable microscope conditions as defined by the Rayleigh criterion and the linear imaging approximation. Figure 7 demonstrates such behavior for a 300 keV instrument with an 11 mrad probe forming semi-angle. For very thin samples, the ADF-STEM image matches well with the linear incoherent imaging model and the Si [211] structure is unresolved. However, for a thicker 20 nm sample (Fig. 7c), the Si [211] structure appears resolved but the "atomic" positions are not in their expected locations (~26% further apart). Figure 7c. shows that these artifacts are most pronounced for realistic sample thicknesses in the range of 10-40 nm and remain for substantially thick specimens (100 nm or more Figure 8 shows the apparent separation of the [211] Si dumbbell for an aberration-corrected ($C_5$=20 mm) 300 keV Titan as the probe size varies. Here this was achieved by varying the size of probe-forming aperture as this provides a hard and unambiguous cutoff for the information limit in the linear imaging approximation. A similar effect could also be achieved by introducing a progressively larger incoherent source size. While for thin specimens (2 nm) the linear imaging approximation holds, in thicker specimens a false dumbbell is present, even when the aperture is reduced below the information limit needed for the true dumbbell spacing. The dumbbells can appear without information transfer beyond the microscopes information limit, however work by Liu & Cowley and Hillyard & Silcox (esp. their Fig 8a) has demonstrated under some conditions it is possible to see Bragg spots in diffractogram beyond the information limit as a result of channeling artifacts[17, 20]. These spots reflect distortions in the image and should not be interpreted as improved resolution. We observe similar results (Supp. Fig. 5)[18].

In general, the false dumbbell spacing is larger than that of the true atom locations. However in thicker samples and small aperture sizes, the dumbbell spacing is reduced, crosses the "correct" spacing as the aperture is increased and continues to increase, reaching a maximum and then decreasing and finally asymptoting to the correct spacing. This is illustrated in the 50 nm curve, where the correct spacings, albeit with reduced contrast, can also be seen for an aperture size that should have been too small to resolve this spacing. The lesson is that even if the dumbbells are resolved at their correct positions, it does not mean that the probe is as small as the dumbbell spacing—here a probe larger than one Angstrom has produced an image with a sub-Angstrom (0.078 nm) spacing. The artifact could be detected by repeating the measurement at a series of different sample thicknesses. If the probe is too large, then at many thicknesses, the dumbbell spacing will be too large as well and will vary with thickness.

Additionally, there is a noticeable polarity of the dumbbell HAADF intensity in the experimental image (Fig. 1). This polarity is a real effect seen in the multislice simulation (Fig. 6) where the intensity of the right column is higher over a range of realistic thicknesses. This asymmetry is reflected in the Si [211] zone axis, where the positions of atoms along one column are shifted along the [211] direction relative to the adjacent column such that symmetry between the left and right column is broken. A linear imaging model fails to predict the polarity of a dumbbell that is seen in experiment and simulation.

The multislice simulations confirm the rather unexpected tight-binding prediction of increased dumbbell spacing and also demonstrate the failure of resolution tests based on the assumption of a simple linear imaging model, or the independent column approximation, at realistic and typical sample thicknesses.

IV. CONCLUSIONS

In summary, we have shown that a simple two-dimensional molecular system captures the key physical trends for fast electron propagation along crystal zone axes, as well as predicting real imaging artifacts found in experimental and simulated ADF-STEM images. When viewing a crystal down a principle zone axis, as is done to obtain atomic images, we've shown that the complexity of the problem can be reduced to textbook simplicity by mapping the propagating beam to the time evolution of a non-stationary state of a 2D-columnar "molecule". As to efficiency, while Bloch plane-waves scale as $O(N^3)$, multislice scales as $O((N\log N)^2)$, but the coupled-columnar approximation scales as $O(N)$, which could reduce the length of some of simulations from days or weeks to minutes or hours for electron propagation through crystals. While imaging of crystals is now possible with sub-Angstrom electron beams produced by a new generation of aberration-corrected microscopes, the propagation of the electron beam can complicate image interpretation. When atomic columns with sufficiently close proximity are observed, the excitation of the resulting 2D molecular orbital's have distinct characteristic signatures in the images that we are able to observe experimentally, and can drastically and predictably change the apparent location of atoms in samples currently used as resolution tests. The shifts in the apparent column spacings suggest caution in directly reading off atomic displacements from ADF-STEM images of grain boundaries and interfaces when atom columns are sufficiently closely spaced to generate molecular orbitals.

ACKNOWLEDGEMENTS

RH thanks Julia A. Mundy for proof reading the manuscript. HLX thanks Judy J. Cha and Earl J. Kirkland for initial code for the radial Schrödinger. RH was supported by Semiconductor Research Corporation. HLX supported by the Energy Materials Center at Cornell, an Energy Frontier Research Center funded by the U.S. Department of Energy, Office of Basic Energy Sciences (DE-SC0001086). Electron microscope facilities supported by the Cornell Center for Materials Research, an NSF MRSEC (NSF DMR-1120296).

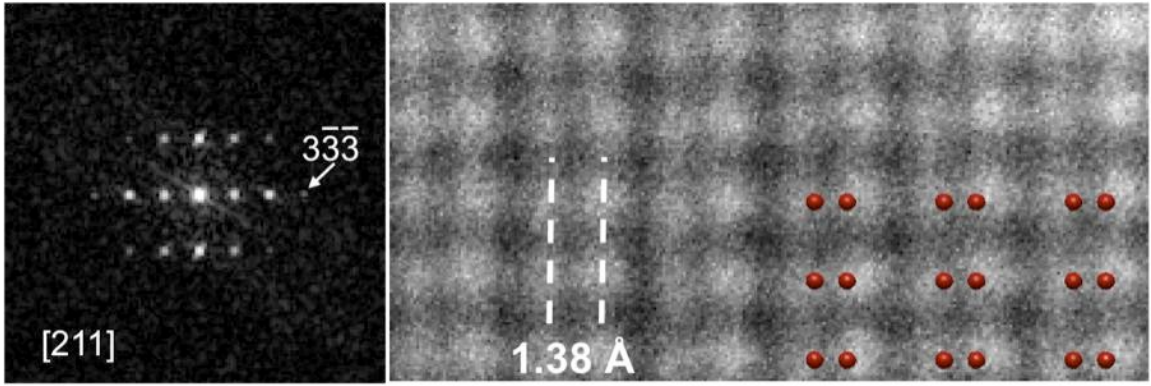

FIG. 1. An ADF-STEM image (8 successively acquired images cross correlated and averaged to increase the SNR) of Si along the [211] zone axis acquired by a 100 KeV aberration-corrected Nion UltraSTEM ($\alpha_{max}$ = 33 mrad, I = 30 pA). Considering the resolution degradation from the chromatic aberration and the incoherent source size, the 0.78 Å spaced dumbbells are not expected to be resolved by this microscope. However, the image shows apparent but unphysical atomic columns with a separation much wider than the actual spacing. Red dots (*lower right*) show the actual atomic positions, which lie closer together by 0.78 Å than the experimental peaks, giving a 'squinted eye' appearance to the composite.

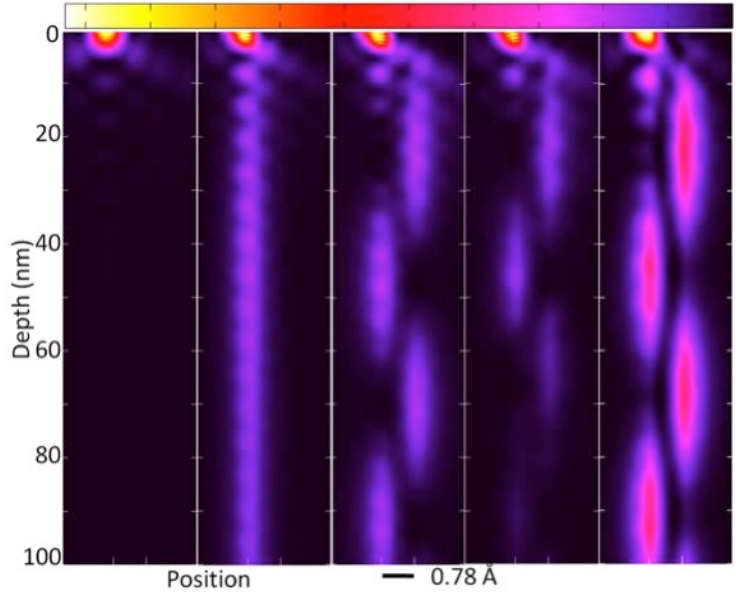

FIG. 2. Cross sectional depth profile of probability for an aberration corrected electron probe (33 mrad, 100 keV) as it propagates a) in free space, b) down a single isolated atomic column, c) down two adjacent isolated columns, d) down two adjacent columns in a full lattice. The atomic columns, atomic spacings, and column spacings are that of the Si [211] zone axis. The electron probe is focused at the entrance surface and positioned just left (0.2 Å) of the atomic columns. The probability density remains localized deep into the sample (over 1000 Å) as it oscillates between atomic columns. The frequency of oscillation is determined by the difference of the eigenenergies of the transverse bonding and anti-bonding states. (a-d) are calculated using the full multislice method. (e) is the analytic tight-binding approach to (c) as described in the text.

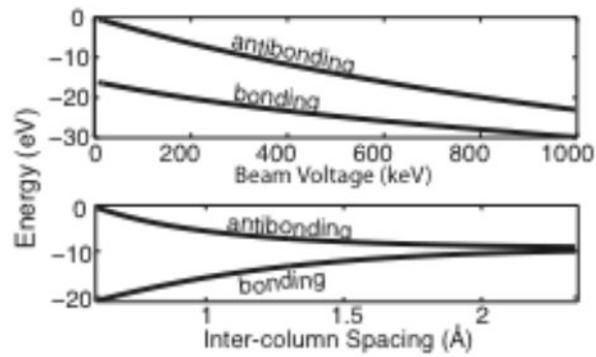

FIG. 3. The eigenenergies of the transverse anti-bonding and bonding states for Si [211] as a function of incident electron energy (top). The plot below shows how the energy levels of the eigenstates split as two atomic columns are brought together.

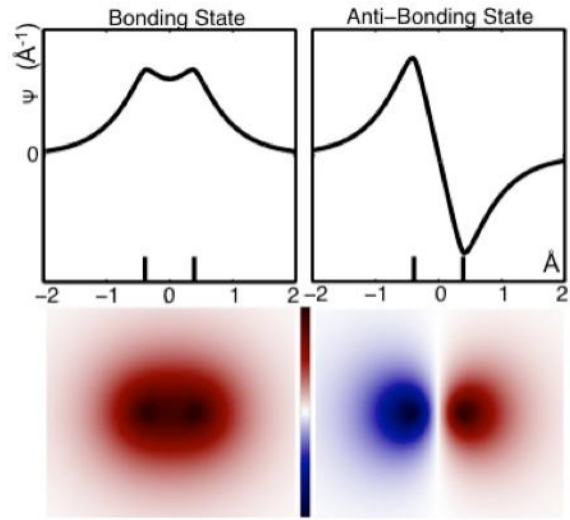

FIG. 4. Bonding (left) and anti-bonding (right) transverse eigenstates of the Si [211] dumbbell structure for a 100 keV electron. Line profiles of the eigenstates are shown (top) with atomic positions marked; corresponding two-dimensional plot shown (bottom).

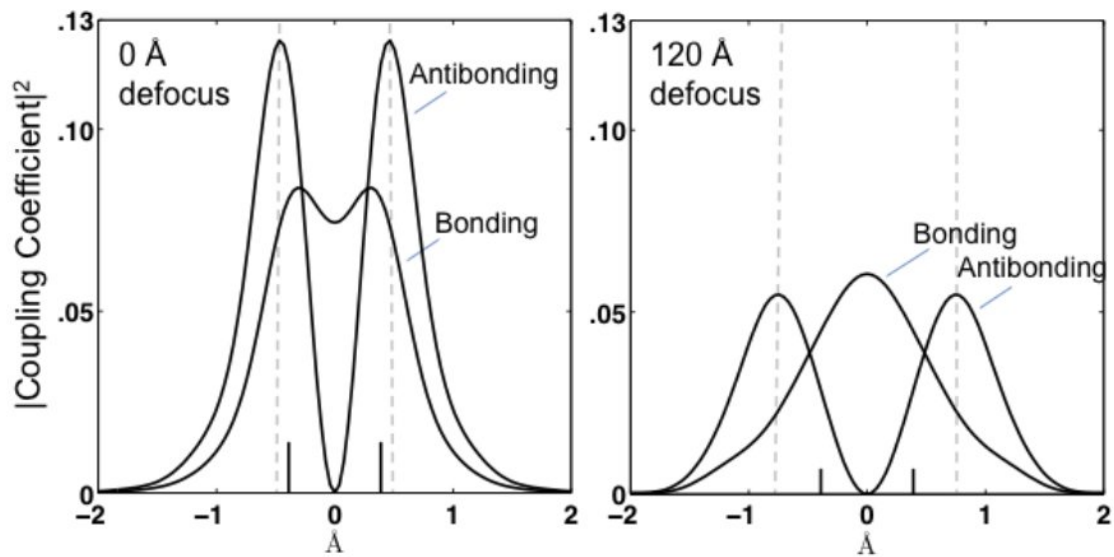

FIG. 5. Squared excitation amplitudes of the bonding and anti-bonding states by a convergent electron probe (100 keV 25 mrad) as a function of the lateral probe position. A probe focused on the surface, defocus =0 Å (left), and a probe focused into the sample, defocus =120 Å (right), are compared. Peak intensities deviate noticeably from atomic column positions as the probe defocus increases.

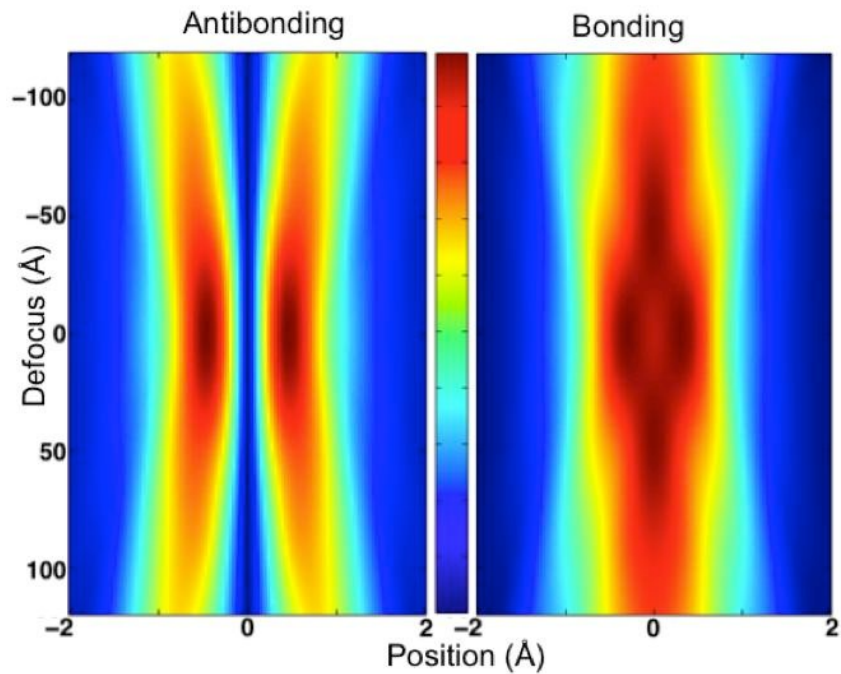

FIG. 6. 2D Map of the excitation amplitudes of the bonding (left) and anti-bonding (right) states by an aberration-free convergent electron probe (100 keV 25 mrad) as a function of both the lateral probe position (x axis) and defocus (y axis). The inter-peak spacing of the anti-bonding excitation is wider than the actual inter-atomic spacing and further widens as the beam is defocused.
18

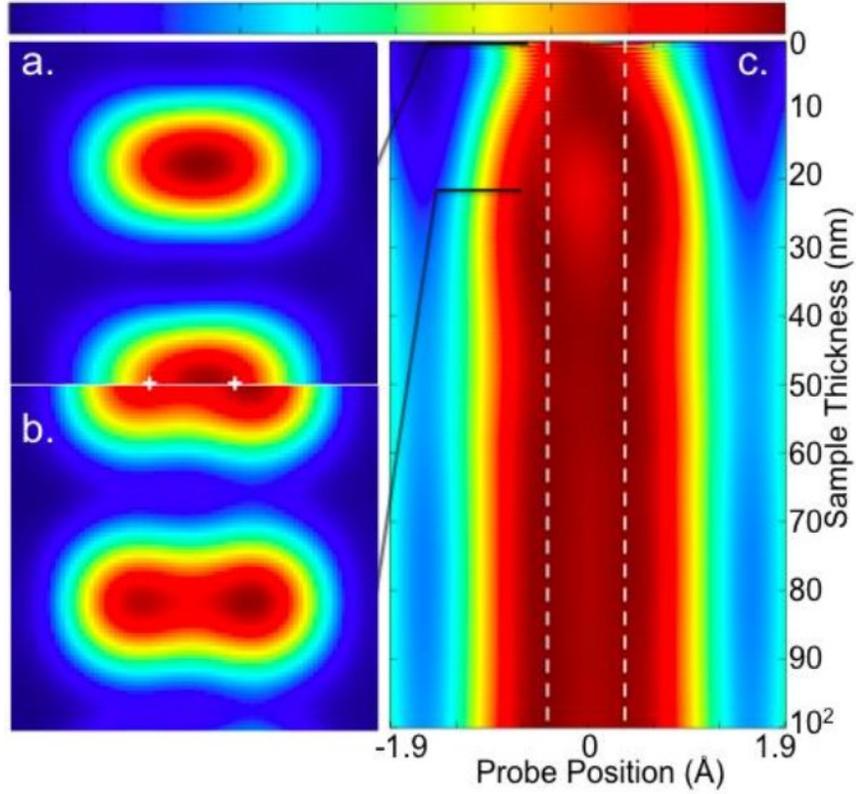

FIG. 7. For a 300 keV electron and 11 mrad probe-forming aperture: Multislice simulation for thin samples (a.) closely matches a simple linear incoherent approximation. However, for thicker samples (b.), the presence of the two 0.78 Å atomic columns (marked in white) become clearly visible despite the 1.09 Å resolving limit of the probe. On the right (c.), line profiles are shown for all thicknesses up to 100nm. Dumbbells are clearly visible around 20-60 nm thicknesses.

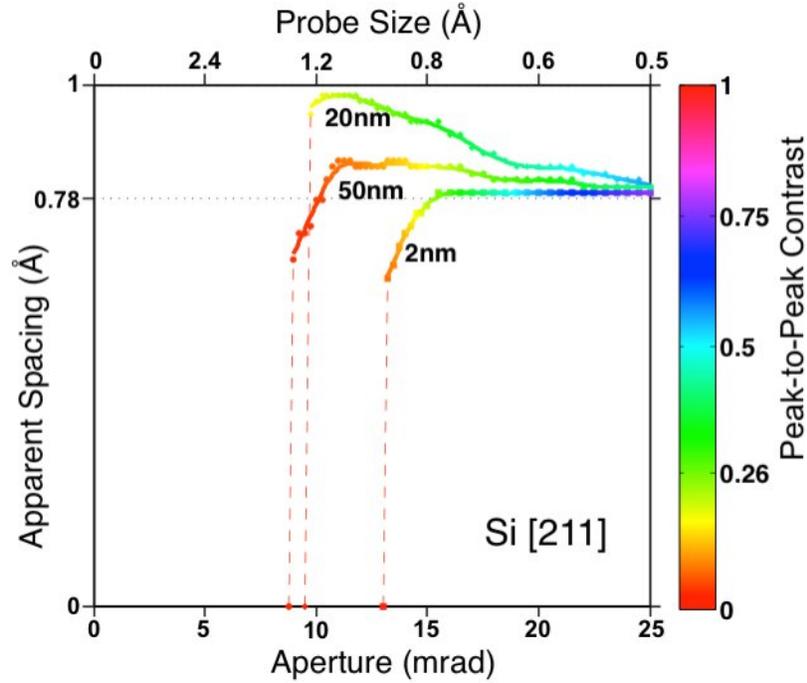

FIG. 8. Apparent separation of the [211] Si dumbbell for an aberration-corrected ($C_5$=20 mm) 300 keV Titan as a function of the probe-forming aperture size. Multislice simulations were ran for 2, 20 and 50 nm thick samples. For thin specimens (2 nm) a linear imaging approximation holds well—for probes smaller than the actual 0.78 Å spacing, dumbbells appear at the correct position. This spacing decreases past the Rayleigh Criterion until it reaches the spacing reaches zero (Sparrow Criterion). However, for thicker specimens, a false dumbbell with an incorrect spacing is present, even when the aperture is reduced below the information limit for the true dumbbell spacing. Correct spacings with reduced contrast can also be seen beyond the transfer limit of the microscope, as shown by the 50nm curve. The probe size was calculated from the Raleigh Criterion.